\pdfoutput=1
\documentclass[12pt, A4paper]{article}
\usepackage [utf8]{inputenc}

\usepackage[shortcuts]{extdash}
\usepackage[nottoc,numbib]{tocbibind}

\usepackage{csquotes}
\usepackage [english]{babel}
\usepackage{graphicx} 
\usepackage[left=1.5cm,right=1.5cm,top=1.5cm,bottom=1.5cm,includefoot,footskip=1.5cm]{geometry}
\usepackage{amsmath}  
\usepackage{cancel}
\usepackage{slashed}
\usepackage[mathscr]{eucal}
\usepackage{rotating,indentfirst,array,varioref}
\usepackage{appendix,marginnote,tikz,pgf,mathtools}
\usepackage{setspace}
\usepackage{misccorr}
\usepackage{siunitx}
\usepackage{amsthm}
\usepackage{amssymb}
\usepackage{wrapfig}
\addto\captionsrussian{
  \renewcommand{\contentsname}%
    {Contents}%
}
\usepackage{hyperref}
\hypersetup{
    colorlinks=true,
    linktoc=all,    
    linkcolor=blue, 
}
\usepackage{titlesec}
\titleformat{\chapter}
  {\normalfont\LARGE\bfseries}{\thechapter}{1em}{}
\titlespacing*{\chapter}{0pt}{3.5ex plus 1ex minus .2ex}{2.3ex plus .2ex}

\usepackage{authblk}

\usepackage[
backend=biber,
sorting=none
]{biblatex}
\newcommand{\pd}{\partial}

\newcommand{\br}[1]{{\overline{#1}}}

\def\<{\left(}
\def\>{\right)}

\begin{document}

\title{Note on large-$p$ limit of $(2,2p+1)$ minimal Liouville gravity and moduli space volumes}

\author{A.A.Artemev $^{1,2,3}$\thanks{artemev.aa@phystech.edu}}

\affil{$^1$ Moscow Institute of Physics and Technology, Institutskii per. 9, 141700,
Dolgoprudny, Russia }

\affil{$^2$Landau Institute for Theoretical Physics, 142432, Chernogolovka, Russia}

\affil{$^3$Skolkovo Institute of Science and Technology, 121205, Moscow, Russia}

\maketitle
\begin{abstract}
In this note we report on some properties of correlation numbers for 2-dimensional Liouville gravity coupled with $(2,2p+1)$ minimal model at large $p$. In the limit $p \to \infty$, for some explicitly known examples in a particular region of parameter space correlation numbers are shown to reduce to Weil-Petersson volumes, analytically continued to imaginary geodesic lengths. This marks another connection of this limit with JT-gravity. We also comment on supposed geometric meaning of the obtained answers outside of this region, in particular, the meaning of the minimal model fusion rules. Another observation is the proportionality of correlation number to the number of conformal blocks when $p$ is big enough compared to parameters of the correlator. This proportionality is valid even without taking the limit.
\end{abstract}
\tableofcontents
\section{Introduction}

There exist several approaches to two-dimensional quantum gravity. Most prominent ones are Witten-Kontsevich TQFT, "discrete" approach via matrix models and Polyakov-Liouville-Distler-Kawai formulation using conformal field theory. A particular model of interest in the last paradigm is minimal Liouville gravity (MLG), where dynamical metric is coupled with CFT minimal model. Coincidence of available results in different approaches leads to a speculation that they all are interconnected with one another; in fact, the essence of (already proven) Witten's conjecture is that the first two approaches are equivalent (for the review of this and related questions see \cite{Dijkgraaf:2018vnm}). Connection between matrix model and MLG is more subtle, since not many calculations are available at the moment on CFT side; generating function for a specific class of correlators on a sphere in MLG is known to coincide with an analogue for matrix model after a certain analytic choice of parameters ("resonance transformations") only up to fourth order \cite{belzam2009}. It is widely believed that the coincidence is exact up to all orders and there is no evidence to suggest otherwise, so we will treat two approaches as equivalent in this note and sometimes loosely refer to certain matrix model correlators as "MLG results" as well. 

Another well-studied model in this realm is Jackiw-Tetelboim (JT) gravity. In this theory, in addition to the metric, we have a scalar field ("dilaton") with no derivative term in the action, which linearly couples to the curvature scalar and forces the path integral over metrics to localize on constant curvature surfaces (there are some nontrivial dynamics on the boundaries of the considered surface if there are such). Such a theory can be argued to coincide with Witten-Kontsevich TQFT in a specific background, where infinitely many couplings are turned on \cite{okuyama2020}. 

Since it was first suggested in \cite{saad2019jt}, connection between JT and $p \to \infty$ limit  of $(2,2p+1)$ minimal Liouville gravity (in the future we will refer to it as "JT limit" or "semiclassical limit" interchangeably) was discussed by several authors. It can be explored by comparing the computable correlators from both sides (see e.g. ref. \cite{mertens2021}), as well as characteristics of corresponding matrix models (for example, see \cite{turiaci2021} or \cite{schiappa}). This work is a certain continuation of these endeavours. 

The note is divided into three parts. First (section \ref{sect1}), we give a quick introduction to the objects of our study --- MLG correlation numbers and JT gravity. The note is not supposed to be self-contained in that regard, so we omit the details, but give the relevant references.  In the second part (section \ref{sect2}), we study $p\to \infty$ behaviour of some correlation numbers in MLG as obtained both from matrix model approach and by field-theoretic methods.  We identify the region in parameter space when familiar expressions for Weil-Petersson volumes for surfaces with geodesic boundaries can be extracted from correlation numbers and provide some explanation on why this region exists. Outside of this region, some properties of these numbers can supposedly also be understood in terms of moduli space geometry; in particular, the fact that correlators vanish when minimal model fusion rules are not satisfied. In a sense, the comparison of JT correlators with results for correlation numbers from matrix model was already performed in more general setting (on the level of generating function and string equations) in \cite{turiaci2021}. However, we find it more illustrative to look at specific expressions. Finally, in section \ref{sect4} we take note of the certain unusual property of correlation numbers that does not require to take the limit $p \to \infty$, but is valid for large enough $p$ in comparison to parameters of the correlator; that is, proportionality of the correlation numbers to number of conformal blocks in minimal model sector. We check this for matrix-model answers for 4- and 5-point functions. The author is not aware of the possible precise explanation of this property in JT limit (apart from a few comments in section \ref{43sec}) or any interpretation in geometrical terms; it would be interesting to understand it in the future.
\section{Preliminaries} \label{sect1}
\subsection{Minimal Liouville gravity} \label{mlgintro}

For a general introduction to discussed approaches to the matter see e.g. \cite{franc1995}, \cite{ginsparg1993lectures}. 

Minimal Liouville gravity is a CFT of total central charge 0 that consists of Liouville CFT, minimal model CFT $\mathcal{M}_{r,q}$ defined by two relatively prime numbers $r,q$ ($r \leq q$) and BRST ghosts:
\begin{equation}
A_{MLG} = \underbrace{\int d^2x\,\left(\frac{1}{4\pi} (\pd_a \varphi)^2 + \mu e^{2b\varphi} \right)}_{A_{Liouv}}+A_{\mathcal{M}_{r,q}}+\underbrace{\frac{1}{\pi}\int d^2x\,\left( C \br{\pd} B + \br{C} \pd \br{B} \right)}_{A_{ghost}}
\end{equation}
 This theory can be derived from the theory of $\mathcal{M}_{r,q}$ CFT minimally coupled to two-dimensional dynamical metric (for that we need additional assumptions on the properties of measure in integral over metrics such as the ones of Distler and Kawai \cite{DISTLER1989509}; results that inspired them to treat two-dimensional gravity this way are due to Polyakov \cite{Polyakov:1987zb} and Knizhnik-Polyakov-Zamolodchikov \cite{Knizhnik:1988ak}).  The remnant Liouville model has a central charge defined by the matter content: requirement of vanishing total central charge (or, equivalently $c_\text{Liouv} = 26-c_\text{M}$) leads to coincidence of Liouville CFT parameter $b$ and  $\beta = \sqrt{r/q}$. In the following only the case of $(r,q) = (2,2p+1)$ will be important to us.

Physical observables in the theory belong to BRST cohomology. The simplest ones of ghost number zero can be built from primary fields of minimal model $\Phi_{1,k+1}$,\, $0 \leq k \leq p-1$ by dressing them with Liouville operators $V_{a} \equiv e^{2 a\varphi}$ such that the total conformal dimension of the resulting operator $U_{1,k+1} \equiv V_a \Phi_{1,k+1}$ is equal to $(1,1)$ (that requires $a = b \frac{k+2}{2}$) and then integrating it over the manifold. One can further dress $U$ with BRST ghosts $C$ to make a $(0,0)$-form $W_{1,k+1} \equiv C \br{C} U_{1,k+1}$ which is a cohomology representative of ghost number 1. Multipoint correlation functions of operators $\int d^2x\,U(x)$ and $W(x)$ (or, more precisely, correlation numbers, since they do not depend on points of insertion of $W$ operators) are of interest to us in this note; ghost number anomaly on the sphere requires the number of $W$ operators in such a correlator to be equal to 3.

An immediate property of this correlator is the so-called "fusion rules": the correlator is equal to zero whenever for its minimal model part in the OPE of corresponding operators $\Phi_{m,n}$ there is no contribution from conformal family of unity. OPE in $(2,2p+1)$ minimal model schematically looks like $[\Phi_{1,k_1+1}] [\Phi_{1,k_2+1}] = \sum \limits_{k= |k_2 - k_1|:2}^{k_1+k_2} [\Phi_{1,k+1}]$, where the field $\Phi_{1,k+1}$ for $k>p-1$ is identified with $\Phi_{1,2p-k}$ and "$:2$" denotes that we change the index of summation by 2 every step. From this property it follows that there are contributions from unit operator (correlator is nonzero) if and only if
\begin{equation}
    \begin{cases}
    k_1 + k_2 + k_3 > k_4,\, \sum k_i \text{ is even;} \\
    k_1 + k_2 + k_3 + k_4 > 2p - 5,\,  \sum k_i \text{ is odd} 
    \end{cases} \label{fusru}
\end{equation}
Three-point correlation numbers can be found just by multiplying results for Liouville CFT, minimal model and ghosts three point-functions \cite{Zamolodchikov:2005fy}; all of these are known exactly. An interesting property is that after certain normalization of the operators $W$ three-point numbers reduce just to fusion constants of underlying CFT minimal model (i.e. are either zero or one); neither in MM CFT or Liouville theory three-point functions are that trivial. When integrated operators $\int d^2x\,U(x)$ are added, situation becomes more difficult; a formal expression as an integrated sum of squares of conformal blocks exists, as always in conformal field theory, but is not convenient for purposes other than numerical evaluation (although Zamolodchikov representation of conformal blocks via series in $q$ allows to do it quite efficiently, at least for four-point correlator, see e.g. \cite{alesh2016}).

In matrix-model approach to $(2,2p+1)$ minimal gravity correlation numbers are derivatives of (the singular part of) the generating functional $\mathcal{Z}(t_0, \dots, t_{p-1})$ in the vicinity of $p$-critical point. This set of relevant deformations of matrix model potential is in one-to-one correspondence with the set of operators $U$ or $W$ \cite{Moore:1991ir}. Calculation of different correlation numbers on the sphere can be streamlined with the help of the so-called "string equation" and is certainly more straightforward than any available field-theoretic calculations. It is, however, somewhat nontrivial to connect two results with each other: generating functional in matrix model $\mathcal{Z}(t_0, \dots, t_{p-1})$ does not coincide with the one of MLG, but they are (conjecturally) related by a certain analytic change of parameters $t = t (\lambda)$. In \cite{belzam2009} the general form of these analytic transformations was proposed.
\subsection{JT gravity and Weil-Petersson volumes}
Jackiw-Tetelboim theory on a surface $M$ with boundary $\pd M$ is defined by an action
\begin{equation}
A_{JT} = - \frac{S_0}{2\pi} \left(\frac{1}{2} \int \limits_{M} \sqrt{g}\, R + \int \limits_{\pd M} \sqrt{h}\, K \right) - \left(\frac{1}{2} \int \limits_M \sqrt{g} \phi (R+2) + \int \limits_{\pd M} \sqrt{h} \phi(K-1) \right)
\end{equation}
where $g$ is the metric on the surface $M$, $h$ --- induced metric on its boundary, $R$ is a Ricci scalar and $K$ external curvature of the boundary. For simplicity we write the action only for the hyperbolic case. The action should also be supplemented with some boundary conditions; usual ones are to fix the lengths of boundary components and value of dilaton $\phi$ on the boundary. Boundary dynamics will not be relevant to our discussion of correlators on a sphere; they are known to reduce to Schwarzian quantum mechanics \cite{saad2019jt}. Surfaces with fluctuating boundaries can be thought of as surfaces with geodesic boundary components with "trumpets" (hyperbolic annuli with one geodesic boundary component and one boundary component of arbitrary shape) glued to them.

For the surface with geodesic boundaries, integrating $\phi$ in the bulk leaves us with "delta-function" $\delta(R+2)$, reducing evaluation of the functional integral to volumes of moduli spaces of hyperbolic metrics. The measure on this space that comes from the functional integral can be argued to coincide with the natural Weil-Petersson (WP) one (i.e. quantum corrections to this behaviour are cancelled). A possible way to derive this is to rewrite hyperbolic JT gravity in the first-order formalism in terms of 2D BF theory with the gauge group $SL(2,\mathbb{R})$ (or, more rigorously, a certain subsemigroup of this, see e.g. \cite{blom2019}) and then to show coincidence of measure for this gauge theory with WP \cite{saad2019jt}. Cancellation of quantum corrections to moduli space measure for BF theory was proved in \cite{Witten:1991we}.

For a survey of results concerning Weil-Petersson volumes see e.g. \cite{do2011moduli}. They can be efficiently calculated using Mirzakhani recursion relation \cite{Mirzakhani:2006fta}. For convenience we list some results for WP volumes of hyperbolic spheres with $n$ geodesic boundaries ($l_i \equiv \frac{L_i}{2\pi}$, where $L_i$ are the boundary lengths) 
\begin{equation}
    V_{0,3} = 1;
\end{equation}
\begin{equation}
    V_{0,4} =  \frac{4\pi^2}{2}\left( \sum \limits_{i=1}^4 l_i^2+ 1 \right); \label{wp04}
\end{equation}
\begin{equation}
    V_{0,5}(L_1, L_2, L_3, L_4,L_5) =  \frac{(2\pi)^4}{8}\left(5 + \sum \limits_{i=1}^5 l_i^4 + 4 \sum \limits_{1 \leq i < j \leq 5} l_i^2 l_j^2 + 6 \sum \limits_{i=1}^5 l_i^2  \right) \label{wp05}
\end{equation}
\section{Weil-Petersson volumes from correlation numbers} \label{sect2}

\subsection{Four-point function from "matrix model"}
We start with the analysis of the $p \to \infty$ limit of the four-point function as obtained from matrix model approach. The "resonance transformations" allowing to obtain this result were first found by Belavin and Zamolodchikov in \cite{belzam2009}; see also \cite{tarn2011} for more explicit expressions in different regions of parameter space. For concreteness let us first consider the case when the sum of parameters $k_i$ (the meaning of them is the same as discussed in section \ref{mlgintro} --- they are parameters of the operators $U$ and $W$ in the correlator) is even; in semiclassical limit one would assume that it should not matter, however, the result for the correlation function in matrix model is actually sensitive to whether it is even or odd (see \cite{belzam2009} for reference). In the future to be more concise we will sometimes refer to these two possibilities as "even" and "odd" sector respectively.

Let us order the parameters of the correlator as $0 \leq k_1 \leq k_2 \leq k_3 \leq k_4 \leq p-1$. Exact formula as obtained in the references is
\begin{equation}
   Z_{k_1 k_2 k_3 k_4} = -F_\theta(-2) + \sum \limits_{i=1}^4 F_\theta(k_i-1) - F_\theta(k_{12|34}) - F_\theta(k_{13|24}) - F_\theta(k_{14|23}) \label{4pfmm}
\end{equation}
where $k_{ij|lm}$ and the function $F_\theta$ are defined as
\begin{equation}
 k_{ij|lm} = \text{min} (k_i + k_j, k_l + k_m) ;\quad   F_\theta(k) = \frac{1}{2} (p-k-1)(p-k-2) \theta(p-2-k)
\end{equation}
There is also the "integral term", but it does not contribute if the fusion rules are satisfied. Consider the case when all operators are "heavy" in the sense that $k$ scales with $p$ when $p \to \infty$ such that $\kappa_i = \frac{k_i}{p}$ stays finite. Then the leading asymptotic for function $F$ is
\begin{equation}
 F_\theta(\kappa p) \approx p^2 \cdot \frac{1}{2} (1-\kappa)^2 \theta(1-\kappa)
\end{equation}
and the first four terms in (\ref{4pfmm}) yield
\begin{equation}
    Z_{k_1 k_2 k_3 k_4} \approx p^2 \cdot \frac{1}{2} \left(\sum \limits_{i=1}^4 (1-\kappa_i)^2 - 1 \right) \label{4pfwp}
\end{equation}
We indeed see that it is proportional to expression in (\ref{wp04}) for  imaginary length parameters $L_i=2\pi i (1-\kappa_i)$. However, for (\ref{4pfwp}) to be the full expression for the correlator the last three terms in (\ref{4pfmm}) should be zero. That means that parameters in corresponding theta-functions should become negative:
\begin{equation}
k_1 + k_2 > p-2 \Rightarrow \kappa_1 + \kappa_2 > 1 \label{cond}
\end{equation}
By our convention, $k_1, k_2$ are less than two other parameters, so if (\ref{cond}) is satisfied, then for all $i\neq j \neq l \neq m$ $k_{ij|lm} > p$. We will see later that (\ref{cond}) is also necessary and sufficient to obtain agreement with analytic continuation of (\ref{wp05}) from five-point correlation function in the large-$p$ limit. In passing we note that from (\ref{cond}) follows the validity of fusion rules both for even and odd values of $\sum k_i$ (see (\ref{fusru})). In the odd sector, the expression for correlator when $k_1 + k_2 > p$ is actually the same (\cite{tarn2011}), so when comparing this correlator in the semiclassical limit with WP volumes "discrete details" are indeed irrelevant. Outside of the region defined by (\ref{cond}) discreteness might actually matter.
\subsection{Five-point function}
Matrix-model result for five-point correlation number is the main result of \cite{tarn2011}; no field-theoretic calculation for it is available at the moment, in contrast with four-point correlator. It will be more convenient for us to use expressions from the appendix of \cite{tarn2011}; we again restrict first to the even sector and order the parameters of correlator according to $0 \leq k_1 \leq k_2 \leq k_3 \leq k_4 \leq k_5 \leq p-1$. Then we obtain
\begin{align}
& Z_{k_1 k_2 k_3 k_4 k_5} = Z^{(I)} + Z^{(J)} + Z^{(1)} + Z^{(2)} \\
& Z^{(1)} = \sum \limits_{i=1}^5 \left(\frac{3p(p+1)}{2} F_\theta(k_i-1) -H_\theta(k_i-2) \right) - 2 \sum \limits_{i<j} F_\theta(k_i-1) F_\theta(k_j-1) - \frac{p(p+1)(5p^2+5p+2)}{8}; \\
& Z^{(2)} = \sum \limits_{i < j} \left(H_\theta (k_{ij}-1) - F_\theta(k_{ij}) \frac{p^2+p}{2} + F_\theta(k_{ij}) \sum \limits_{l \neq i,j} F_\theta(k_l - 1) \right) -  \sum \limits_{i<j<l} H_\theta(k_{ijl}) - \sum \limits_{i,j,l,m} F_\theta(k_{ij}) F_\theta(k_{lm} ); \\
& Z^{(I)} = \sum \limits_n \frac{1}{8}(2k_n - k -2) (2k_n - k -4) (2p-3-k)(2p-5-k)\, \theta (2k_n - k -6); \\
& Z^{(J)} = \sum \limits_{i<j} \left(H(k-k_{ij}) - \frac{(k-2k_{ij})(k-2k_{ij}+2)(2p-3-k)(2p-5-k)}{8} \theta(2k_{ij}-k-2) \right) \times \\ \nonumber
& \times \theta (p-1-k+k_{ij}) \theta(p-1-k_{ij})
\end{align} 
Here we have introduced new notations
\begin{equation}
H_\theta(k) = \frac{1}{2} F_\theta(k) F_\theta(k+2);\quad k = \sum \limits_{i=1}^5 k_i,\,k_{ij} = k_i + k_j,\,k_{lmn} = k_l + k_m + k_n
\end{equation}
($H$ is the same as $H_\theta$ without the theta-function). The answer is much more bulky than the one for four-point correlator, but when (\ref{cond}) is satisfied, in the semiclassical limit only $Z^{(1)}$ is nonzero. Indeed, arguments of $H_\theta$ and $F_\theta$ in $Z^{(2)}$ are larger than $p$, and since they both contain a theta-function $\theta(p-1-k)$, they become zero; same theta-function is present for every term in the sum in $Z^{(J)}$; finally, since $k_n < p~~\forall n$ and $k> \frac{5p}{2}$ when (\ref{cond}) is valid, theta-function $\theta(2k_n - k -6)$ in $Z^{(I)}$ is zero as well. The semiclassical limit of $Z^{(1)}$ is now easy to take; we have
\begin{equation}
    H_\theta(\kappa p) \approx p^4 \cdot \frac{1}{8} (1-\kappa)^4 \theta(1-\kappa)
\end{equation}
and thus, we obtain
\begin{equation}
    Z^{(1)} \approx -\frac{p^4}{8} \left(5 + \sum \limits_{i=1}^5 (1- \kappa_i)^4 + 4 \sum \limits_{1 \leq i < j \leq 5} (1- \kappa_i)^2 (1- \kappa_j)^2 - 6 \sum \limits_{i=1}^5 (1- \kappa_i)^2  \right)
\end{equation}
in accordance with the expression (\ref{wp05}) up to a prefactor. Again (\ref{cond}) guarantees the validity of the fusion rules for both even and the odd sectors. In the odd case, the only difference in the expression for correlator is that $Z^{(I)}$ and $Z^{(J)}$ are equal to zero identically; so, again, if $\sum k_i$ is odd or even does not matter for our purpose here.
\subsection{Four-point function from "higher equations of motion"} \label{hem}
For the case of four-point number, there is another approach to calculating the correlator in continuum formulation, using methods of conformal field theory (specifically, the so-called "higher equations of motion" (HEM) \cite{highereoms2004} that allow to express Liouville dressing operator for one of the operators in terms of derivative of degenerate field with respect to conformal dimension; after such a substitution OPEs of fields in the theory are calculated relatively straightforwardly, if one takes into account BRST-symmetry of the theory). The expression can be written as
\begin{equation}
    Z^{HEM}_{k_1 k_2 k_3 k_4} = -(1+k_1)(p+k_1+\frac{3}{2}) + \sum \limits_{i=2}^4 \sum \limits_{s=-k_1:2}^{k_1} |p-s-k_i - \frac{1}{2}| \label{hemeq}
\end{equation}
The derivation can be found in \cite{Belavin:2005jy}. A certain subtlety is that this calculation assumes a specific matter theory instead of minimal model. For this theory, the name "generalized minimal model" was (somewhat confusingly) used in different sources. Perhaps it would be more correct to call it "$c<1$ Liouville theory", which was shown to exist and be a consistent CFT in \cite{clessthan12015}. What is considered is a theory with generic (irrational) central charge $c<1$ where the structure constants are specific solutions of recursion relations following from conformal bootstrap. These structure constants are such that when calculated for degenerate dimensions they coincide with the ones from minimal model almost always  \cite{Zamolodchikov:2005fy}. But, apart from degenerate fields in the Kac table, operators with generic dimension also can be considered in this theory, unlike the minimal model; the expression for the four-point correlator (\ref{hemeq}) that we use is supposed to be valid for 1 degenerate and 3 generic operator insertions (i.e. $k_i$ for $i \neq 1$ are supposed to be non-integer). This answer for four-point number is known to be consistent \cite{belzam2009} with matrix model when the assumptions in the derivation (which can be reformulated as an assumption on the number of conformal blocks in minimal model sector) are valid; the corrections to these answers when these are violated are also understood \cite{alesh2016}. 

We think it is interesting to compare this result with (\ref{wp04}) independently, as this could be thought of as another check of that JT limit is independent of discrete details of the MLG correlator (it would mean that it does not matter if parameters $k_i$ are integer or generic). Also, this is somewhat nontrivial since expression (\ref{hemeq}) is piecewise linear in $k_i,\, i \neq 1$ yet (\ref{wp04}) is quadratic. The reason for coincidence, however, is that this piecewise linear function has a lot of derivative discontinuities in our $p\to \infty$ limit and becomes almost quadratic in the limit. Let us derive (\ref{wp04}) from this expression; we assume the validity of (\ref{cond}) for this. Dropping subleading terms, rescaling $s = p x,\,x \in (-\kappa_1, \kappa_1)$ and rewriting $\sum \limits_s$ as an integral $\frac{p}{2} \int dx$, we have
\begin{equation}
    \frac{Z^{HEM}_{k_1 k_2 k_3 k_4}}{p^2} \approx -\kappa_1 - \kappa_1^2 + \frac{1}{2} \sum \limits_{i=2}^4 \int \limits_{-\kappa_1}^{\kappa_1} dx\,|1-x-\kappa_i| = -\kappa_1 - \kappa_1^2 + \sum \limits_{i=2}^4 \frac{\kappa_1^2 + (1-\kappa_i)^2}{2} = \frac{1}{2} \left(\sum \limits_{i=1}^4 (1-\kappa_i)^2 - 1 \right)
\end{equation}
We see that it coincides with analytically continued to imaginary lengths expression (\ref{wp04}) up to prefactor with no assumptions on parameters of the correlator other than (\ref{cond}).
\subsection{Explanation of the coincidence} \label{34sec}
It is long known in Liouville field theory (see e.g. \cite{seibergnotes}) that heavy operators from physical spectrum (for real Liouville momenta $P = \sqrt{\Delta - \frac{Q^2}{4}}$) in semiclassical limit are "non-local" and correspond to macroscopic "holes" in the surface, geometry of which is described by Liouville field. In accordance with this understanding, in  \cite{mertens2021} MLG fixed-boundary-length bulk one-point function on the disk was computed and found to correspond to "trumpet" geometries (with one geodesic and one fluctuating boundary) in JT limit; length of geodesic boundary corresponding to operator of momentum $P$ is $L = 2\pi \cdot 2b \cdot P$. 

However, dressing operators for minimal model primaries that we consider have imaginary Liouville momenta. As also argued in \cite{mertens2021}, they correspond to conical defects where $P$ defines the angle deficit. Let us give an intuitive explanation for why imaginary length geodesics are connected with conical defects. Consider some closed geodesic of length $2\pi \epsilon$ on a hyperbolic surface; due to well-known collar lemma from hyperbolic geometry in some neighbourhood of it one can choose coordinates $t, \alpha \in [0,2\pi)$ such as the metric becomes one of hyperbolic cylinders:
\begin{equation}
    ds^2 = \frac{dt^2}{t^2+\epsilon^2} + (t^2 + \epsilon^2) d\alpha^2,
\end{equation}
where $\alpha$ is a (rescaled) proper length coordinate along the considered geodesic and $t$ is more or less distance to this geodesic from the considered point. Now analytically continue $\epsilon^2$ to negative values $\epsilon = i \tilde{\epsilon}$ and restrict to the region $t>\tilde{\epsilon}$; by a change of coordinates $t = \tilde{\epsilon} \cosh r$ we obtain
\begin{equation}
    ds^2 = dr^2 + \tilde{\epsilon}^2 \sinh^2 r\, d\alpha^2,
\end{equation}
which is the hyperbolic metric in the vicinity of conical defect of angle deficit $2\pi (1-\tilde{\epsilon})$. Thus, what we calculate in the semiclassical limit is something like "volumes of moduli spaces of constant curvature surfaces with conical defects of fixed deficit angle". We stress that they are not necessarily hyperbolic: conical defects contribute to the RHS in Gauss-Bonnet theorem (we write it for sphere topology)
\begin{equation}
R \cdot \text{Area}(\Sigma) \sim 2 - \sum \limits_{i=1}^n \kappa_i, \label{gaussbonnet}
\end{equation}
thus, the sign of curvature for uniformizing metric is determined by defect angles. Conical angle and parameter of the field in the correlator are related by
\begin{equation}
  \tilde{\epsilon} = 1 -\kappa  \label{condefkappa}
\end{equation}
To obtain the answers for the volumes of such moduli spaces one might also try to just use formulas for moduli spaces of surfaces with geodesic boundaries with imaginary length parameters.  However, one should be careful --- not all parts of the construction of moduli space withstand this "analytic continuation"; one of the particular facts that causes trouble is that for surfaces with conical defects in general case there are homotopy classes for which there is no geodesic representative (see e.g. \cite{Witten:2020wvy} for explanation of this).

To our knowledge, such moduli spaces (and whether there are some naturally defined volumes associated to them) are not well understood rigorously in general. A particular mathematical result \cite{Tan2004GeneralizationsOM} suggests that if all the defects are "sharp" enough (in our notations that means $\kappa_i > 1/2~~\forall i$), then analytic continuation works. In fact, the corresponding case was already studied in the physics literature for JT gravity (e.g. \cite{Witten:2020wvy}, \cite{Maxfield:2020ale}), where explicit computations confirmed that moduli space volumes for surfaces with both boundaries and "sharp" defects are given by analytic continuations in some of the arguments of expressions for surfaces with boundaries only (in particular, formula for four-point number (\ref{4pfwp}) was obtained). However, investigating semiclassical limit of minimal string correlators apparently allows to obtain more information: firstly, we found that correlation numbers in semiclassical limit coincide with analytically continued expressions for surfaces with boundaries even when a slightly weaker (than requirements for all defects to be "sharp") condition (\ref{cond}) holds. Secondly, we expect that outside of the region of "sharp" defects  semiclassical limits of MLG correlators also have some geometrical meaning (as a particular example see section \ref{sectroyanovreg} below).

Additional terms appear when (\ref{cond}) is not valid: for example, for four-point function if only for first two defects $\kappa_1 + \kappa_2 < 1$, we get an additional contribution to the answer of the form $- (1 - \kappa_1 - \kappa_2)^2$, which is reminiscent of what we would get for conical defect with angle deficit $\kappa_1 + \kappa_2$. This is suggestive of that the possibility of "merging" of defects affects the moduli space metric (such explanation was already proposed in \cite{turiaci2021}). In this case such merging is allowed, by which we mean that there exists a continuous family of constant curvature metrics which interpolate between the case of two conical singularities and one of the combined deficit angle \cite{xuwenrafe2020}. In terms of moduli space it could mean the appearance of some singularities which have the structure of moduli spaces with less number of cones. It is interesting to see if this can be reformulated more rigorously and in a less speculative manner.

We conclude discussion of the connection with WP volumes with a comment on properties of correlation numbers at finite $p$. In \cite{mertens2021}, the so-called "p-deformed" Weil-Petersson volumes were extracted from multi-boundary amplitudes in minimal string theory. Such a deformation would be one possible way to come from JT to minimal gravity correlators. However, if we analytically continue these deformed volumes to imaginary geodesic lengths, finite $p$ corrections of \cite{mertens2021} do not seem to coincide with the corresponding ones in the correlation numbers we discussed, at least naively. Perhaps agreement could be achieved by redefining how Liouville momenta is connected with conical defect angle, adding some subleading in $1/p$ terms to the formula (\ref{condefkappa}).
\subsection{Fusion rules and "Troyanov region"} \label{sectroyanovreg}
Now we give a small comment on one most evident property of investigated correlators outside of the region where (\ref{cond}) is valid; namely, the fact that correlators become zero when the fusion rules (\ref{fusru}) are not satisfied. For an arbitrary $n$-point function in the even sector that means that
\begin{equation}
\sum \limits_{i=1}^{n-1} \kappa_i < \kappa_n \label{fusion}
\end{equation}
Because $\kappa_n < 1$, from Gauss-Bonnet theorem (\ref{gaussbonnet}) we would necessarily have curvature $R>0$ in this case if constant curvature metrics with such conical singularities existed. 

Geometric meaning of (\ref{fusion}) is actually well known: the region in parameter space where the opposite inequality $\sum \limits_{i=1}^{n-1} \kappa_i > \kappa_n$ is valid is usually referred to as "Troyanov region". The condition that parameters $\kappa_i$ belong to Troyanov region is known to be necessary for the existence of a metric of constant positive curvature with given conical angles on a sphere (for a proof of this see e.g. \cite{mazzeo2015teichmuller}). If parameters do not belong to Troyanov region, such metrics do not exist and thus  for moduli space volume (however it is defined) it would be natural to become zero, consistently with fusion rules for corresponding correlator.
\section{Another property of the correlators at large $p$} \label{sect4}
In this section we analyze the matrix model results for multipoint correlators in another region of parameter space and prove that there the only nontrivial dependence on relative values of $k_i$ is in the proportionality to the number of conformal blocks. This region is defined by the property that sum of any two numbers $k_i + k_j,\,i \neq j$ is less than $p$ for the four-point function and the same is valid for the sum of any three numbers for five-point function.
\subsection{Four-point function}
We start with the four-point correlator; assume that $\sum k_i$ is even for simplicity. 

The number of conformal blocks can be easily calculated; it depends on relation between $k_4 + k_1$ and $k_2+k_3$ (the ordering is as before). If $k_4 + k_1 < k_2 + k_3$, all the conformal families that appear in the OPE of $\Phi_{1, k_1+1} \Phi_{1, k_4+1}$ also appear for $\Phi_{1, k_2+1} \Phi_{1, k_3+1}$; their number is equal to the number of conformal blocks and is $1+k_1$. In the other case, active channels in the OPE are $k \in (k_4 - k_1, k_4 - k_1 + 2, \dots,\,k_2+k_3)$   and the number of them is $1 + \frac{k_2 + k_3 + k_1 - k_4}{2}$. Using the expression for four-point number in the form given in \cite{tarn2011}, for the case that we consider ($p> k_3 + k_4$) we have
\begin{equation}
Z_{k_1 k_2 k_3 k_4} = 
\begin{cases}
(1+k_1)(2p-3-k),\,k_{14}<k_{23} \\
(1 + \frac{k_2 + k_3 + k_1 - k_4}{2})(2p-3-k),\,k_{14} \geq k_{23} \\
\end{cases} \label{4pfconfbl}
\end{equation}
as announced before. 

It can be noted that the remaining factor $(2p-3-k)$ has a simple meaning expressed via the "number of screenings" $n$ in the Liouville part of the correlator; it is $\sim (n+1)$ up to factor of two and a sign. In fact, in this domain formula for the correlator (in the case $k_{14}<k_{23}$) is in accordance (up to overall normalization independent of correlator parameters that we ignore) with the one derived by Fateev and Litvinov from Coulomb integrals \cite{fatlit2008} obtained by "analytic continuation" from the case of integer number of screenings. As their derivation, similarly to the one based on HEM, supposes that three of the fields in the correlator are non-degenerate, it has the same shortcoming in that it coincides with matrix model result only when there are $1+k_1$ conformal blocks in the decomposition. However, to "correct" their formula, it turns out to be enough to replace one the prefactor with the true number of conformal blocks. In the domain that we consider HEM formula (\ref{hemeq}) also reduces to the first line of (\ref{4pfconfbl}). 
\subsection{Five-point function}
It is not so simple in this case to derive the general formula for the number of conformal blocks, because there are a lot of possible cases for the relations among $k_i$. However, it is easy to establish the similar "factorized" form of five-point correlator obtained from the matrix model. In this case we will need an additional assumption that for any $i \neq j \neq l$ $k_{ijl}<p$ as well. We start with the formula as obtained in the last section of \cite{tarn2011}:
\begin{align}
Z_{k_1 k_2 k_3 k_4 k_5} &= \frac{1}{8} \left(4 \sum_i k_i^2 - k^2 -2k-8 - \sum \limits_{m<n}G_3 (m,n) (k-2k_{mn})(k-2k_{mn}+2) \right) (2p-3-k)(2p-5-k) \nonumber + \\
&+ \sum \limits_{i<j<l} H(k_{ijl}) \theta(k_{ijl} - p ) \theta(k - k_{ijl} - p) - \sum \limits_{i,j,l,m} F(k_{ij}) F(k_{lm}) \theta(k_{ij}-p) \theta(k_{lm}-p) + \nonumber \\
&+\sum \limits_n \frac{1}{8}(2k_n - k -2) (2k_n - k -4) (2p-3-k)(2p-5-k)\, \theta (2k_n - k -6),
\end{align}
where $G_3$ is defined by
\begin{equation}
G_3(m,n) = \theta(k_{mn} - p) + \theta(2k_{mn}-k-2) \theta (p-1 - k + k_{mn}) \theta (p-1-k_{mn})
\end{equation}
In our domain $G_3$ is simplified to $\theta(2k_{mn} - k - 2)$; the terms in the second line are all zero due to negative arguments of theta-functions, and the other two terms contain the same factor $(2p-3-k)(2p-5-k)$ (which is, again, expressed as $\sim (n+1)(n+2)$ in terms of number of screenings; compare with the answer for 5-point number by Fateev and Litvinov \cite{fatlit2008}). Other than this factor we have
\begin{align}
\frac{Z_{k_1 k_2 k_3 k_4 k_5}}{(2p-3-k)(2p-5-k)}&=\frac{1}{8} \left(4 \sum_i k_i^2 - k^2 -2k-8 - \sum \limits_{m<n}\theta(2k_{mn} - k - 2) (k-2k_{mn})(k-2k_{mn}+2) \right. \nonumber \\
&+ \left.  \sum \limits_n (2k_n - k -2) (2k_n - k -4) \right)
\end{align}
One can check numerically for some examples that this expression indeed gives the number of conformal blocks (with negative sign) in the decomposition of the five-point function in the even sector.
\subsection{Some comments} \label{43sec}
The conditions that we used to derive our results are in a sense opposite to what was discussed in the first part when comparing to WP volumes: while there the coincidence was obtained when no two defects are allowed to merge in a one with larger deficit angle, here we consider the case when any two (or even any three, for the five-point case) defects have the sum of deficit angle less than $2\pi$ and thus can merge. For such a case, we see in the moduli space volumes some kind of "large" $sl(2)$ representation (that is defined by number $k_i = \kappa_i \cdot p $) corresponding to each defect $\kappa_i$; multiplying these representations and counting the number of trivial summands in the result, we get the moduli space volume up to the factor that depends only on sum of all angle deficits, which is a surprisingly simple property (in fact, in $p \to \infty$ limit this factor reduces to $(2 - \sum \kappa_i)^\#$, which is just the RHS of Gauss-Bonnet theorem (\ref{gaussbonnet})). It would be interesting to understand what this means geometrically and why the "mergeability" is relevant here.

The fact that the correlation number "counts" the number of conformal blocks is reminiscent of Wilson loop correlators in Chern-Simons theory, which is well known to be connected with two-dimensional CFTs (see e.g. \cite{cmp/1104178138}). Perhaps this might be seen as one explanation of the discussed property: Chern-Simons theory at large level $k \to \infty$ reduces to BF-theory that for the case of gauge group $SL(2,\mathbb{R})$, as mentioned before, is a presumably equivalent description of JT gravity. Appearance of $SL(2,\mathbb{R})$ as Chern-Simons/BF gauge group is expected since the corresponding current algebra is known to emerge in the effective gravity theory induced by conformal matter (\cite{Bershadsky:1989mf}, \cite{Polyakov:1987zb}, \cite{Knizhnik:1988ak}). This argument needs to be worked out in more detail, however; we leave this for a future work together with examining the validity of the discussed property for higher correlation numbers.  

\section{Conclusions and discussion}
The discussed properties inspire several possible directions for further studies. First one is "physical": analysis of the semiclassical limit of correlation numbers could give some physical intuition on a structure of MLG CFT correlators. In particular, it would be interesting to see a simple way of deriving a property described in section \ref{sect4} directly from CFT approach and to understand whether it is applicable to other multipoint correlators. The second is "geometrical", i.e. concerns the cone surface moduli spaces volumes: we propose a certain "physical" definition of such volumes via limits of matrix model (or CFT) correlators, that has some expected properties (for example, Troyanov/fusion property discussed in section \ref{sectroyanovreg}). These answers coincide with analytically continued WP volumes for the case of "sharp" defects, which was well understood in earlier studies both from geometrical and JT gravity points of view. However, we believe that more precise mathematical understanding of obtained expressions in the region of parameter space when the defects are not "sharp" is also an important problem. The peculiarities that appear there (such as "contributions from merged conical defects") seem quite interesting to interpret in a geometrical language.
\section{Acknowledgements}
The author is grateful to Konstantin Aleshkin and Alexey Litvinov for valuable discussions. He also thanks Ilya Kochergin for careful reading of the text, Thomas Mertens for correspondence on the matter and Sylvain Ribault for useful clarifications on the contents of reference \cite{clessthan12015}. This work was supported by the Russian Science Foundation grant (project no. 18-12-00439).
\printbibliography

@article{mertens2021,
    author = "Mertens, Thomas G. and Turiaci, Gustavo J.",
    title = "{Liouville quantum gravity -- holography, JT and matrices}",
    eprint = "2006.07072",
    archivePrefix = "arXiv",
    primaryClass = "hep-th",
    doi = "10.1007/JHEP01(2021)073",
    journal = "JHEP",
    volume = "01",
    pages = "073",
    year = "2021"
}

@article{cmp/1104178138,
author = {Edward Witten},
title = {{Quantum field theory and the Jones polynomial}},
volume = {121},
journal = {Communications in Mathematical Physics},
number = {3},
publisher = {Springer},
pages = {351 -- 399},
year = {1989}
}

@article{Moore:1991ir,
    author = "Moore, Gregory W. and Seiberg, Nathan and Staudacher, Matthias",
    title = "{From loops to states in 2-D quantum gravity}",
    reportNumber = "RU-91-11, YCTP-P11-91",
    doi = "10.1016/0550-3213(91)90548-C",
    journal = "Nucl. Phys. B",
    volume = "362",
    pages = "665--709",
    year = "1991"
}

@article{Dijkgraaf:2018vnm,
    author = "Dijkgraaf, Robbert and Witten, Edward",
    title = "{Developments in Topological Gravity}",
    eprint = "1804.03275",
    archivePrefix = "arXiv",
    primaryClass = "hep-th",
    doi = "10.1142/S0217751X18300296",
    journal = "Int. J. Mod. Phys. A",
    volume = "33",
    number = "30",
    pages = "1830029",
    year = "2018"
}

@misc{ginsparg1993lectures,
      title={Lectures on 2D gravity and 2D string theory (TASI 1992)}, 
      author={P. Ginsparg and Gregory Moore},
      year={1993},
      eprint={hep-th/9304011},
      archivePrefix={arXiv},
      primaryClass={hep-th}
}

@article{alesh2016,
    author = "Aleshkin, Konstantin and Belavin, Vladimir",
    title = "{On the construction of the correlation numbers in Minimal Liouville Gravity}",
    eprint = "1610.01558",
    archivePrefix = "arXiv",
    primaryClass = "hep-th",
    doi = "10.1007/JHEP11(2016)142",
    journal = "JHEP",
    volume = "11",
    pages = "142",
    year = "2016"
}

@article{DISTLER1989509,
title = {Conformal field theory and 2D quantum gravity},
journal = {Nuclear Physics B},
volume = {321},
number = {2},
pages = {509-527},
year = {1989},
issn = {0550-3213},
doi = {https://doi.org/10.1016/0550-3213(89)90354-4},
url = {https://www.sciencedirect.com/science/article/pii/0550321389903544},
author = {Jacques Distler and Hikaru Kawai}
}

@article{franc1995,
   title={2D gravity and random matrices},
   volume={254},
   ISSN={0370-1573},
   url={http://dx.doi.org/10.1016/0370-1573(94)00084-G},
   DOI={10.1016/0370-1573(94)00084-g},
   number={1–2},
   journal={Physics Reports},
   publisher={Elsevier BV},
   author={Francesco, P.Di and Ginsparg, P. and Zinn-Justin, J.},
   year={1995},
   month={Mar},
   pages={1–133} }

@article{Bershadsky:1989mf,
    author = "Bershadsky, Michael and Ooguri, Hirosi",
    title = "{Hidden SL(n) Symmetry in Conformal Field Theories}",
    reportNumber = "IASSNS-HEP-89-09",
    doi = "10.1007/BF02124331",
    journal = "Commun. Math. Phys.",
    volume = "126",
    pages = "49",
    year = "1989"
}

@article{Witten:2020wvy,
    author = "Witten, Edward",
    title = "{Matrix Models and Deformations of JT Gravity}",
    eprint = "2006.13414",
    archivePrefix = "arXiv",
    primaryClass = "hep-th",
    doi = "10.1098/rspa.2020.0582",
    journal = "Proc. Roy. Soc. Lond. A",
    volume = "476",
    number = "2244",
    pages = "20200582",
    year = "2020"
}

@article{Maxfield:2020ale,
    author = "Maxfield, Henry and Turiaci, Gustavo J.",
    title = "{The path integral of 3D gravity near extremality; or, JT gravity with defects as a matrix integral}",
    eprint = "2006.11317",
    archivePrefix = "arXiv",
    primaryClass = "hep-th",
    doi = "10.1007/JHEP01(2021)118",
    journal = "JHEP",
    volume = "01",
    pages = "118",
    year = "2021"
}

@article{Polyakov:1987zb,
    author = "Polyakov, Alexander M.",
    title = "{Quantum Gravity in Two-Dimensions}",
    reportNumber = "Print-87-1052 (LANDAU)",
    doi = "10.1142/S0217732387001130",
    journal = "Mod. Phys. Lett. A",
    volume = "2",
    pages = "893",
    year = "1987"
}

@misc{schiappa,
  doi = {10.48550/ARXIV.2108.11409},
  
  url = {https://arxiv.org/abs/2108.11409},
  
  author = {Gregori, Paolo and Schiappa, Ricardo},
  
  title = {From Minimal Strings towards Jackiw-Teitelboim Gravity: On their Resurgence, Resonance, and Black Holes},
  
  publisher = {arXiv},
  
  year = {2021},
  
  copyright = {arXiv.org perpetual, non-exclusive license}
}

@article{Knizhnik:1988ak,
    author = "Knizhnik, V. G. and Polyakov, Alexander M. and Zamolodchikov, A. B.",
    editor = "Khalatnikov, I. M. and Mineev, V. P.",
    title = "{Fractal Structure of 2D Quantum Gravity}",
    reportNumber = "PRINT-88-0812 (LANDAU)",
    doi = "10.1142/S0217732388000982",
    journal = "Mod. Phys. Lett. A",
    volume = "3",
    pages = "819",
    year = "1988"
}

@article{Zamolodchikov:2005fy,
    author = "Zamolodchikov, Alexei B.",
    title = "{Three-point function in the minimal Liouville gravity}",
    eprint = "hep-th/0505063",
    archivePrefix = "arXiv",
    reportNumber = "LPM-04-10",
    doi = "10.1007/s11232-005-0003-3",
    journal = "Theor. Math. Phys.",
    volume = "142",
    pages = "183--196",
    year = "2005"
}

@misc{do2011moduli,
      title={Moduli spaces of hyperbolic surfaces and their Weil-Petersson volumes}, 
      author={Norman Do},
      year={2011},
      eprint={1103.4674},
      archivePrefix={arXiv},
      primaryClass={math.GT}
}

@article{turiaci2021,
    author = "Turiaci, Gustavo J. and Usatyuk, Mykhaylo and Weng, Wayne W.",
    title = "{2D dilaton-gravity, deformations of the minimal string, and matrix models}",
    eprint = "2011.06038",
    archivePrefix = "arXiv",
    primaryClass = "hep-th",
    doi = "10.1088/1361-6382/ac25df",
    journal = "Class. Quant. Grav.",
    volume = "38",
    number = "20",
    pages = "204001",
    year = "2021"
}

@misc{saad2019jt,
      title={JT gravity as a matrix integral}, 
      author={Phil Saad and Stephen H. Shenker and Douglas Stanford},
      year={2019},
      eprint={1903.11115},
      archivePrefix={arXiv},
      primaryClass={hep-th}
}

@article{Belavin:2005jy,
    author = "Belavin, A. A. and Zamolodchikov, A. B.",
    title = "{Integrals over moduli spaces, ground ring, and four-point function in minimal Liouville gravity}",
    doi = "10.1007/s11232-006-0075-8",
    journal = "Theor. Math. Phys.",
    volume = "147",
    pages = "729--754",
    year = "2006"
}

@article{seibergnotes,
    author = {Seiberg, Nathan},
    title = "{Notes on Quantum Liouville Theory and Quantum Gravity}",
    journal = {Progress of Theoretical Physics Supplement},
    volume = {102},
    pages = {319-349},
    year = {1990},
    month = {03},
    issn = {0375-9687},
    doi = {10.1143/PTP.102.319},
    url = {https://doi.org/10.1143/PTP.102.319},
    eprint = {https://academic.oup.com/ptps/article-pdf/doi/10.1143/PTP.102.319/5376238/102-319.pdf},
}

@article{clessthan12015,
    author = "Ribault, Sylvain and Santachiara, Raoul",
    title = "{Liouville theory with a central charge less than one}",
    eprint = "1503.02067",
    archivePrefix = "arXiv",
    primaryClass = "hep-th",
    doi = "10.1007/JHEP08(2015)109",
    journal = "JHEP",
    volume = "08",
    pages = "109",
    year = "2015"
}

@article{highereoms2004,
    author = "Zamolodchikov, A.",
    editor = "Belavin, A. and Corrigan, Edward",
    title = "{Higher equations of motion in Liouville field theory}",
    eprint = "hep-th/0312279",
    archivePrefix = "arXiv",
    reportNumber = "PM-03-34",
    doi = "10.1142/S0217751X04020592",
    journal = "Int. J. Mod. Phys. A",
    volume = "19S2",
    pages = "510--523",
    year = "2004"
}

@article{tarn2011,
    author = "Tarnopolsky, G.",
    title = "{Five-point Correlation Numbers in One-Matrix Model}",
    eprint = "0912.4971",
    archivePrefix = "arXiv",
    primaryClass = "hep-th",
    doi = "10.1088/1751-8113/44/32/325401",
    journal = "J. Phys. A",
    volume = "44",
    pages = "325401",
    year = "2011"
}

@article{Tan2004GeneralizationsOM,
  title={Generalizations of McShane's identity to hyperbolic cone-surfaces},
  author={Ser Peow Tan and Yan Loi Wong and Ying Zhang},
  journal={Journal of Differential Geometry},
  year={2004},
  volume={72},
  pages={73-112}
}

@article{Witten:1991we,
    author = "Witten, Edward",
    title = "{On quantum gauge theories in two-dimensions}",
    doi = "10.1007/BF02100009",
    journal = "Commun. Math. Phys.",
    volume = "141",
    pages = "153--209",
    year = "1991"
}

@article{blom2019,
    author = "Blommaert, Andreas and Mertens, Thomas G. and Verschelde, Henri",
    title = "{Fine Structure of Jackiw-Teitelboim Quantum Gravity}",
    eprint = "1812.00918",
    archivePrefix = "arXiv",
    primaryClass = "hep-th",
    doi = "10.1007/JHEP09(2019)066",
    journal = "JHEP",
    volume = "09",
    pages = "066",
    year = "2019"
}

@article{okuyama2020,
    author = "Okuyama, Kazumi and Sakai, Kazuhiro",
    title = "{JT gravity, KdV equations and macroscopic loop operators}",
    eprint = "1911.01659",
    archivePrefix = "arXiv",
    primaryClass = "hep-th",
    doi = "10.1007/JHEP01(2020)156",
    journal = "JHEP",
    volume = "01",
    pages = "156",
    year = "2020"
}

@article{xuwenrafe2020,
   title={Conical metrics on Riemann surfaces, I: The compactified configuration space and regularity},
   volume={24},
   ISSN={1465-3060},
   url={http://dx.doi.org/10.2140/gt.2020.24.309},
   DOI={10.2140/gt.2020.24.309},
   number={1},
   journal={Geometry I\& Topology},
   publisher={Mathematical Sciences Publishers},
   author={Mazzeo, Rafe and Zhu, Xuwen},
   year={2020},
   month={Mar},
   pages={309–372}
}

@article{fatlit2008,
    author = "Fateev, V. A. and Litvinov, A. V.",
    title = "{Multipoint correlation functions in Liouville field theory and minimal Liouville gravity}",
    eprint = "0707.1664",
    archivePrefix = "arXiv",
    primaryClass = "hep-th",
    doi = "10.1007/s11232-008-0038-3",
    journal = "Theor. Math. Phys.",
    volume = "154",
    pages = "454--472",
    year = "2008"
}

@misc{mazzeo2015teichmuller,
      title={Teichm\"uller theory for conic surfaces}, 
      author={Rafe Mazzeo and Hartmut Weiss},
      year={2015},
      eprint={1509.07608},
      archivePrefix={arXiv},
      primaryClass={math.DG}
}

@article{belzam2009,
    author = "Belavin, A. A. and Zamolodchikov, A. B.",
    title = "{On Correlation Numbers in 2D Minimal Gravity and Matrix Models}",
    eprint = "0811.0450",
    archivePrefix = "arXiv",
    primaryClass = "hep-th",
    reportNumber = "RUNHETC-2008-24",
    doi = "10.1088/1751-8113/42/30/304004",
    journal = "J. Phys. A",
    volume = "42",
    pages = "304004",
    year = "2009"
}

@article{Mirzakhani:2006fta,
    author = "Mirzakhani, Maryam",
    title = "{Simple geodesics and Weil-Petersson volumes of moduli spaces of bordered Riemann surfaces}",
    doi = "10.1007/s00222-006-0013-2",
    journal = "Invent. Math.",
    volume = "167",
    number = "1",
    pages = "179--222",
    year = "2006"
}
\end{document}